\documentclass[twocolumn,pre,aps,flushbottom,showpacs]{revtex4-1}
\usepackage{xspace,amsmath,amsfonts,amsthm,amssymb,amsbsy,graphicx,color}

\begin{document}

\newcommand{\ms}[1]{\mbox{\scriptsize #1}}
\newcommand{\mbs}[1]{\mbox{\scriptsize $\mathbf{#1}$}}

\title{Quantum effects improve the energy efficiency of feedback control}

\author{Jordan M.~Horowitz$^1$}
\author{Kurt Jacobs$^{1,2}$}
\affiliation{$^1$Department of Physics, University of Massachusetts at Boston, Boston, MA 02125, USA} 
\affiliation{$^2$Hearne Institute for Theoretical Physics, Louisiana State University, Baton Rouge, LA 70803, USA} 

\begin{abstract}  
The laws of thermodynamics apply equally well to quantum systems as to classical systems, and because of this quantum effects do not change the fundamental thermodynamic efficiency of isothermal refrigerators or engines. We show that, despite this fact, quantum mechanics permits measurement-based feedback control protocols that are more thermodynamically efficient than their classical counterparts. As part of our analysis we perform a detailed accounting of the thermodynamics of unitary feedback control and elucidate the sources of inefficiency in measurement-based and coherent feedback.  
\end{abstract}


\maketitle 

\section{Introduction}

Feedback control is a process in which information obtained about a system is used to control its future evolution, often with the goal of pushing it into a target configuration. Effective control not only requires the ability to direct the motion of the system, but also the capability to remove unwanted noise. This latter task entails reducing the uncertainty about the system by extracting entropy. The information utilized by feedback may be obtained by making explicit measurements, or it may be acquired in a fully coherent manner by correlating the system with another~\cite{Wiseman94, Lloyd00, Hamerly12, Hamerly13, Nurdin09, Jacobs13x}. In fact, the coherent form of feedback is more general, as it subsumes measurement-based feedback as a special case, and as a result coherent feedback can sometimes perform better~\cite{Hamerly12, Hamerly13, Nurdin09, Jacobs13x}. The choice of which form of feedback to use depends on the situation, and both forms have been realized experimentally~\cite{Koch10, Sayrin11, Teufel11, Vijay12, Kerckhoff12}.

Arguably, the most important thermodynamic cost of a process is the energy required, that is the work supplied that must be lost as heat.  
Here, we consider the energy cost of quantum feedback control via unitary operations (``unitary feedback''). The purpose of the feedback is to reduce the entropy of the system by as much as possible, given the constraints. We consider the energy cost of the feedback process, both when the protocol is implemented using explicit measurements, and when the same protocol is implemented in a fully coherent way. 

For unitary feedback control we first show that the energy cost can be characterized by a thermodynamically-consistent efficiency parameter. Simply put, we know from Landauer's principal~\cite{Leff,Esposito2011} that to reduce the von Neuman entropy of a finite quantum system by $\Delta H<0$, a controller must eventually dump the unwanted entropy in a thermal reservoir at ambient temperature $T$. This requires transferring a minimum heat $Q_{\rm min}=-kT \Delta H$ (with $k$ Boltzmann's constant) to the thermal reservoir. By the first law of thermodynamics, $Q_{\rm min}$ represents the minimum energy  that must be supplied by the controller. This suggests that when a feedback process reduces the entropy by $\Delta H$ and dumps $Q$ heat into the surroundings, a well-motivated identification of energetic efficiency is
\begin{equation}\label{eq:eff}
   \varepsilon = \frac{Q_{\ms{min}}}{Q} = -\frac{kT\Delta H}{Q}  \leq 1,
\end{equation}
which takes the form of a standard thermodynamic efficiency given as the ratio of the desired output, $\Delta H$, divided by the required input $Q$~\cite{Diana2013,Zulkowski2013}.
In the following, we perform a careful accounting of the energetics of coherent and measurement-based feedback control, validating Eq.~(\ref{eq:eff}) and elucidating the sources of inefficiency in the control process. 

We show here that the energy efficiency of feedback control implemented by using a measurement (measurement-based feedback) can be increased when the measurement is performed in a basis in which the density matrix of the system is not diagonal. Since this is impossible in classical systems, it is a purely quantum effect. Moreover, it is already known that measuring a single qubit in a basis that is unbiased with respect to the eigenbasis of its density matrix provides the maximum reduction in the average entropy conditioned on the measurement~\cite{rapidP}. Thus, we find that the maximum reduction in entropy coincides with the maximum energetic efficiency.

We begin our discussion in Sec.~\ref{sec:feedCont} with a light introduction to quantum feedback control.  The thermodynamic analysis is carried out in Sec.~\ref{sec:2law}, where we elucidate the inefficiencies of both coherent and measurement-based feedback control.
With those tools in hand, we move onto Sec.~\ref{sec:advantage} where we show how quantum effects can improve the energetic efficiency of feedback control for a simple two-state model. 

%%%%%%%%%%%%%%%%%%%%%%%%%%%%%%%%%%%%%

\section{Quantum feedback control}\label{sec:feedCont}

The basic scenario of feedback control is a quantum system we wish to manipulate,  whose initial state is given by the density matrix $\rho$, and a separate quantum system, or auxiliary, prepared independently in state $\chi$. Feedback control proceeds by applying a sequence of unitary operations to the joint system that shifts any unwanted entropy from the system to the auxiliary.
These operations do work on the system, the energy for which is supplied by a reversible work source. Together, the auxiliary and the work source are the resources for feedback control, and this pair is often called the controller~\cite{Nurdin09, Hamerly12}.

Coherent feedback, in particular, is implemented by applying a joint unitary $U_{\rm coh}$ to the system and auxiliary taking them to the state $r_{\rm coh}$, with system marginal $\rho_{\rm coh}$ and auxiliary marginal $\chi_{\rm coh}$.
If the change in the entropy $H(\rho)=-{\rm Tr}[\rho\ln\rho]$ of the system is negative, that is $\Delta H_S\equiv H(\rho_{\rm coh})-H(\rho)\le 0$, then the entropy of the auxiliary must have increased to compensate, $\Delta H_A\ge -\Delta H_S$ [see Eq.~(\ref{eq:cohEnt}) below]~\cite{Lloyd1989}.
We complete the process by returning the auxiliary to its initial state through an isothermal process, depositing this additional entropy $\Delta H_A$ into the thermal reservoir as heat $Q$. This heat represents the energy that must be supplied during the control process: it is due to work $W$ performed by the reversible work source or may also be supplied by the change in energy of the system $\Delta E_S$.
Together $Q=W-\Delta E_S$ is the energetic cost that we analyze in detail below. 

Measurement-based feedback is a subclass of coherent feedback, where the measurement and feedback steps are separate and distinct~\cite{Jacobs13x}.
Specifically, the system and auxiliary are first correlated via a unitary interaction, $U_{\rm meas}$. The useful information about the system is typically correlated with a preselected basis $\{|m\rangle\}$ of the auxiliary, which we call the measurement basis.
Thus, after the interaction the joint density matrix can be written in the form~\cite{Jacobs2009}
\begin{equation}
  r_{\rm meas}= \sum_m p_m \rho_m \otimes |m\rangle \langle m| + \sum_{m\not=m^\prime} \sigma_{mm^\prime} \otimes |m\rangle \langle m^\prime| , 
\end{equation}
where $p_m$ are probabilities.
The density matrices $\rho_m$ are related to the initial density matrix $\rho$ by $\rho_m=\Pi_m\rho \Pi^\dag_m/p_m$, with $p_m={\rm Tr}[\Pi^\dag_m \Pi_m\rho]$ for some set of operators $\{\Pi_m\}$ that satisfy $\sum_m\Pi_m^\dag \Pi_m=I$.
Each $\rho_m$ represents the state of the system given that the result of the measurement on the auxiliary is $|m\rangle$.
In the special case when the process is chosen to extract information only about  a specific observable $X$ without causing any additional change to the system, then the $\{\Pi_m\}$ must be positive operators that satisfy $[\Pi_m,X]=0$, for all $m$~\cite{Wiseman10}.
Feedback is carried out by applying a different unitary $U_m$ to each $\rho_m$, which can be accomplished by applying the joint unitary $U_{\rm fb}=\sum_m U_m\otimes |m\rangle\langle m|$.
Overall, this is simply a special case of coherent feedback with unitary $U_{\rm fb}U_{\rm meas}$.
Alternatively, we can make the measurement \emph{explicit} by performing a measurement on the auxiliary in the basis $\{|m\rangle\}$ after the application of $U_{\rm meas}$.
Even when we use an explicit measurement in the feedback process, we do not need to apply a projector to describe it; it is enough to decohere the auxiliary in the measurement basis, which means eliminating the relevant off-diagonal elements of $r_{\rm meas}$ to give 
\begin{equation}
  r_{\rm dec} = \sum_m p_m \rho_m \otimes |m\rangle \langle m|.
\end{equation}
Again feedback is carried out by applying the joint unitary $U_{\rm fb}=\sum_mU_m\otimes |m\rangle\langle m|$, resulting in the state $r_{\rm fb}=\sum_m p_mU_m\rho_mU^\dag_m\otimes |m\rangle\langle m|$.
From now on, we will distinguish such \emph{explicit measurement-based feedback} where we decohere the auxiliary after the measurement unitary from that of \emph{coherent measurement-based feedback} where the measurement and feedback processes are implemented separately, but coherently.

%%%%%%%%%%%%%%%%%%%%%%%%%%%%%%%%%%%%%

\section{The second law and thermodynamic efficiency}\label{sec:2law}

We now proceed with a thermodynamic analysis of coherent and explicit measurement-based feedback.  
Our results follow from the well established second law-like inequality for the manipulation of mesoscopic quantum systems in contact with a thermal reservoir at temperature $T$,
\begin{equation}\label{eq:2Law}
\Delta_{\ms{i}}S=k\Delta H+Q/T\ge 0 . 
\end{equation}
This relation may be established both for quantum and classical systems~\footnote{For classical systems the von Neumann entropy is replaced by the Shannon entropy.} as a consequence of the microscopic dynamics~\cite{Hasegawa2010, Takara2010,Esposito2010,Sagawa2013,Reeb2013} or as a general thermodynamic result~\cite{Esposito2011}.
Here, $Q$ is the heat flow into the thermal  reservoir, and we have identified the irreversible entropy production $\Delta_{\ms{i}}S$.
Any process that saturates the inequality in Eq.~(\ref{eq:2Law}) is said to be reversible.
On the other hand, the deviation from equality ($\Delta_{\ms{i}}S > 0$) is a measure of the irreversibility of the process, and is equal to the heat dissipated (irretrievably lost) to the thermal reservoir.
When the process is rapid enough for the dynamics to be unitary, the process is necessarily reversible, since no heat can be transferred to the reservoir, $Q=0$, and the entropy is conserved, $\Delta H=0$.
This is true for coherent quantum devices such as mesoscopic superconducting circuits and nano-mechanical resonators~\cite{Palomaki2013, Kirchmair2013}.
 
 \subsection{Coherent feedback}
 
 The initial phase of coherent feedback control is a unitary interaction.
During this interaction the joint entropy of the system and auxiliary is conserved, $H(r_{\rm coh})=H(r)$, which may be expressed in terms of the change in the auxiliary entropy $\Delta H_A=H(\chi_{\rm coh})-H(\chi)$ and the change in the system entropy $\Delta H_S=H(\rho_{\rm coh})-H(\rho)$ as
 \begin{equation}\label{eq:cohEnt}
 \Delta H_A=-\Delta H_S+I(r_{\rm coh}),
 \end{equation}
 by introducing the quantum mutual information~\cite{Lloyd1989, Vedral2002} 
 \begin{equation}\label{eq:I}
 I(r_{\rm coh})=H(\rho_{\rm coh})+H(\chi_{\rm coh})-H(r_{\rm coh})\ge 0.
 \end{equation}
 The positivity of the mutual information implies that when the system entropy is reduced, there must be a greater increase in the auxiliary, $\Delta H_A\ge -\Delta H_S$~\cite{Lloyd1989}.
 After the feedback step, we return the auxiliary to its initial state.
 This reset operation is performed without any further interaction with the system, since we want to leave the system where we have prepared it.  
 Thus, the heat flow for this process can be deduced from Eq.~(\ref{eq:2Law}) applied solely to the reduced density matrix of the auxiliary, leading to
 \begin{equation}\label{eq:feedCont2}
 Q\ge kT \Delta H_A.
 \end{equation} 
 Substituting in Eq.~(\ref{eq:cohEnt}), we find
 \begin{equation}\label{eq:feedCont}
 Q\ge -kT\Delta H_S+kTI(r_{\rm coh}).
 \end{equation}
 This is our primary result regarding the energetic requirements of coherent feedback control.
 Since $I\ge 0$, we see that in order to reduce the system entropy by $\Delta H_S$, there is a minimum energetic cost in the form of $Q_{\rm min}=-kT\Delta H_S$ heat transferred to the reservoir.
 By the first law of thermodynamics, $Q=W-\Delta E_S\ge 0$, this energy must be supplied either as work $W$ by the work source or by the system itself $\Delta E_S$ (recall that the auxiliary undergoes a cycle, $\Delta E_A=0$). 
We can therefore characterize the energetic efficiency as the factor by which the energy supplied $Q$ exceeds the minimum required $Q_{\rm min}=-kT\Delta H_S$: $\varepsilon=-kT\Delta H_S/Q$ as in Eq.~(\ref{eq:eff}).

 If we assume that we can perform the reset of the auxiliary optimally, we saturate the bound in Eq.~(\ref{eq:feedCont2}), leading to an optimal heat flow
 \begin{equation}\label{eq:opt}
 Q_{\rm opt}= k T \Delta H_A.
 \end{equation}
In this optimal case,
 \begin{equation}\label{eq:eff2}
 \varepsilon=-\frac{\Delta H_S}{\Delta H_A}=\frac{|\Delta H_S|}{|\Delta H_S|+I(r_{\rm coh})}\le 1.
 \end{equation}
Here, the efficiency is determined by how faithfully entropy is transferred from the system to the auxiliary.  Notably, even when the reset operation is performed optimally, feedback control may still be inefficient due to any residual correlations $I(r_{\rm coh})$ between the system and the auxiliary after the unitary interaction.
During the unitary interaction, work was supplied to change the system energy, auxiliary free energy by $\Delta {\mathcal F}_A$, and to form the correlations $I(r_{\rm coh})$.
 During the reset, the  free energy $\Delta {\mathcal F}_A$ supplied to the auxiliary can be extracted back as work; however, the free energy stored in the correlations cannot be recovered, since the reset operation acts only on the auxiliary.
 Instead the correlations are dissipated.
To see this, we apply Eq.~(\ref{eq:2Law}) to the joint system during an optimal reset of the auxiliary, in which the joint density matrix changes from $r_{\rm coh}\to \rho_{\rm coh}\otimes\chi$:
\begin{equation}
\Delta_{\ms{i}} S=H(\rho_{\rm coh}\otimes\chi)-H(r_{\rm coh})+Q_{\rm opt}= kI(r_{\rm coh})\ge 0,
\end{equation}
after substituting in Eqs.~(\ref{eq:cohEnt}) and (\ref{eq:opt}).
Thus, the process is inherently irreversible, since the residual information $I(r_{\rm coh})$ is necessarily dissipated away.
This highlights a key point: any irreversibility during the control process leads to energetic inefficiencies. 
 
 \subsection{Explicit measurement-based feedback}
 
For explicit measurement-based feedback the situation is similar.  Recall that this kind of feedback has three stages: i) establishing the correlation with the auxiliary system,  which we will refer to as the \textit{measurement step}; ii) decohering the auxiliary; and iii) applying the feedback.   
 
 We begin with the measurement step by applying the unitary interaction $U_{\rm meas}$ establishing correlations as in Eq.~(\ref{eq:feedCont2}):
 \begin{equation}\label{eq:mbMeas}
 \Delta H_A^{\rm meas}=\Delta H_S^{\rm meas}-I(r_{\rm meas}). 
 \end{equation}
Here the superscript indicates the stage over which the entropy change is evaluated, for example $\Delta H_A^{\rm meas}=H(\chi_{\rm meas})-H(\chi)$, where $\chi_{\rm meas}$ is the reduced density matrix for the auxiliary at the end of the measurement step. 
After the measurement unitary, the auxiliary is allowed to decohere in the measurement basis $\{|m\rangle\}$. This decoherence can be accomplished by coupling the auxiliary to a thermal reservoir for a time much shorter than the thermal relaxation time, but longer than the decoherence time~\cite{Zurek03}, so long as we choose the measurement basis $\{|m\rangle\}$ to be the energy basis~\footnote{If the measurement basis is not the energy basis, we can always unitarily rotate the Hamiltonian so this is true, then interact the system with the thermal reservoir, and finally rotate the Hamiltonian back. This will add no new inefficiencies, since the rotations are reversible.}. This is not an unreasonable assumption, in light of the ubiquity of thermal reservoirs and the desire for the measurement outcomes to be stable against decoherence.
This allows us to treat the thermodynamics of the decoherence step $r_{\rm meas}\to r_{\rm dec}=\sum_m p_m\rho_m\otimes|m\rangle\langle m|$ using Eq.~(\ref{eq:2Law}):
noting that since by construction the populations of the auxiliary energy eignstates (here $\{|m\rangle\}$) do not change, no heat can flow into the thermal reservoir and Eq.~(\ref{eq:2Law}) implies
\begin{equation}\label{eq:mbDec}
H(r_{\rm dec})-H(r_{\rm meas})=\Delta H_A^{\rm dec}-(I(r_{\rm dec})-I(r_{\rm meas}))\ge 0.
\end{equation}
The result is that the auxiliary entropy increases, while irreversibly destroying correlations.
Such an irreversible process will always lower the energetic efficiency, allowing us to conclude that explicit measurement-based feedback will always have a lower efficiency than its coherent cousin.
After the decoherence step, we apply the feedback unitary $U_{\rm fb}=\sum_m U_m\otimes |m\rangle\langle m|$, which changes the system, but leaves the auxiliary untouched.
Thus, entropy conservation during this step takes the form
\begin{equation}\label{eq:mbFb}
\Delta H_S^{\rm fb}-(I(r_{\rm fb})-I(r_{\rm dec}))=0.
\end{equation}
Finally, we reset the auxiliary back to its initial state, which, just as in Eq.~(\ref{eq:feedCont2}), requires heat
\begin{equation}\label{eq:mbReset}
Q\ge kT\Delta H_A=kT(\Delta H_A^{\rm meas}+\Delta H_A^{\rm dec}),
\end{equation}
where $\Delta H_A=H(\chi_{\rm dec})-H(\chi)$ is the total change in the entropy of the auxiliary over the course of measurement, decoherence, and feedback.
Combining, Eqs.~(\ref{eq:mbMeas}), (\ref{eq:mbDec}), (\ref{eq:mbFb}), and (\ref{eq:mbReset}), we find the energy required to lower the entropy of the system by $\Delta H_S=\Delta H_S^{\rm meas}+\Delta H_S^{\rm fb}$ is bounded by
\begin{equation}
Q\ge -kT\Delta H_S+kTI(r_{\rm fb}),
\end{equation}
analogous to Eq.~(\ref{eq:feedCont}).
Again, $I(r_{\rm fb})$ represents the residual correlations not utilized during feedback.
This leads to an efficiency measure of the form in Eq.~(\ref{eq:eff}), or when the auxiliary is optimally reset 
\begin{equation}\label{eq:effMb}
\varepsilon=-\frac{\Delta H_S}{\Delta H_A}\le \frac{|\Delta H_S|}{|\Delta H_S|+I(r_{\rm fb})}.
\end{equation}
in analogy to Eq.~(\ref{eq:eff2}), where the inequality stems from the fact that the decoherence step is irreversible, thus reducing the efficiency.
It is worth noting that a similar conclusion can be deduced for isothermal measurement-based feedback processes within the framework developed by Sagawa and Ueda~\cite{Sagawa2008,Sagawa2009}. 
In addition, for classical systems, the authors of Ref.~\cite{Diana2013} carried through a similar analysis on the energetic efficiency of finite-time erasing (entropy reduction) of classical bits and arrived at consistent conclusions.
 
%%%%%%%%%%%%%%%%%%%%%%%%%%%%%%%%%%%%%

 \section{Quantum advantage}\label{sec:advantage}
 
 We now apply this general analysis to a specific example and show that a measurement that involves quantum coherences can improve the efficiency of feedback control over one that can be performed classically. Our example consists of a single-qubit system and a single-qubit auxiliary.  
 The system and auxiliary begin in equilibrium at temperature $T$, with Boltzmann density matrices  represented on the Bloch sphere as $\rho=(I-\alpha\sigma_z)/2$ and $\chi=(I-\lambda \sigma_z)/2$ with $0\le \alpha,\lambda\le 1$ and $\vec\sigma=(\sigma_x,\sigma_y,\sigma_z)$ the vector of Pauli matrices: for a given temperature, $\alpha$ and $\lambda$ are determined by the respective energy level splittings of the qubits. Using feedback control, we will reduce the entropy of the system $H(\rho)=h(\alpha)\equiv-(1+\alpha)/2\ln[(1+\alpha)/2]-(1-\alpha)/2\ln[(1-\alpha)/2]$.
This requires that the auxiliary be initially more pure than the system, $\lambda \ge \alpha$, or equivalently have a lower entropy.

\begin{figure}[tb] 
\centering
\includegraphics[scale=.55]{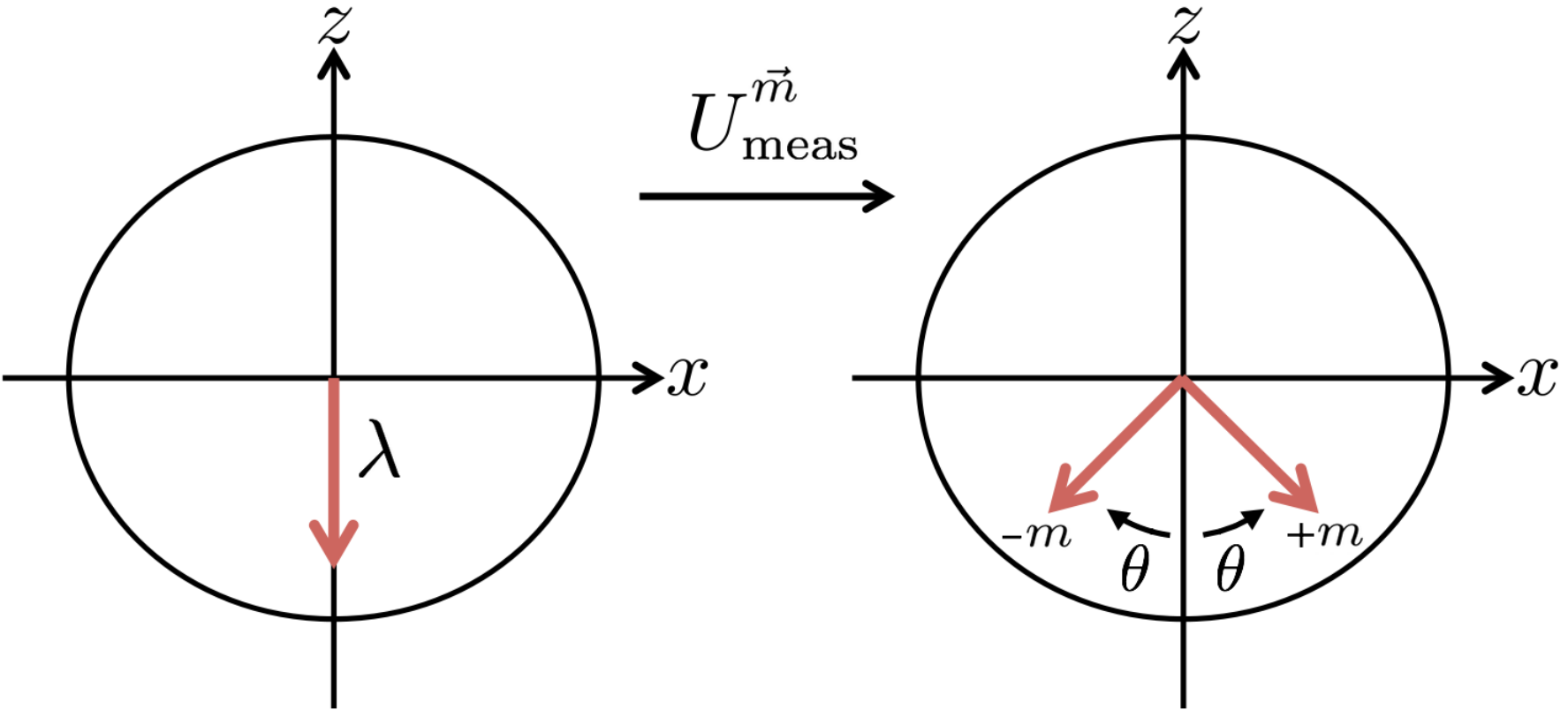}
\caption{Illustration on the Bloch sphere of the conditional rotations of the auxiliary initially in a mixed state with Bloch vector $-\lambda {\hat z}$.
When the system is initially prepared in the $|+ \vec{m}\rangle$ ($|- \vec{m}\rangle$) eigenstate, the auxiliary is conditionally rotated under $U^{\vec m}_{\rm meas}$ by an angle $\theta$ counterclockwise (clockwise) in the $xz$-plane.}
\label{fig:Bloch}
\end{figure}

To perform a measurement of the system in a given basis, the controller must apply a joint unitary that correlates that basis with the auxiliary. 
For our auxiliary, initially diagonal in the $z$-basis, we can measure $\sigma_{\vec m}={\vec m}\cdot \vec\sigma$  with unit vector $\vec{m}$ by applying the unitary $U^{\vec m}_{\rm meas}=e^{-i\theta V_{\vec m}}$ generated by $V_{\vec m}= \sigma_{\vec m}\otimes \sigma_y$.
The state of the auxiliary is rotated on the Bloch sphere about the $y$-axis in opposite directions for each of the eigenstates of $\sigma_{\vec m}$ by an angle $\theta$, illustrated in Fig.~\ref{fig:Bloch}.
The information is now recorded in the auxiliary's $x$-basis: an ideal projective measurement of $\sigma_x$ will obtain the maximum information about $\sigma_{\vec m}$.
When the auxiliary is initially pure ($\lambda=1$) and $\theta=\pi/2$, the measurement is said to be perfect, since projecting the auxiliary on to one of its $\sigma_x$ eigenstates, projects the system onto a pure state. In general, a measurement may not be perfect for two reasons.
First, $\theta <\pi/2$ in which case the measurement is described as being weak.  
In this sense $\theta$ is a measure of the ``strength'' of the measurement.
Second, the auxiliary state can be initially mixed, $\lambda<1$, in which case the measurement is referred to as inefficient~\cite{JacobsSteck06}. 

\begin{figure}
\includegraphics[scale=.55]{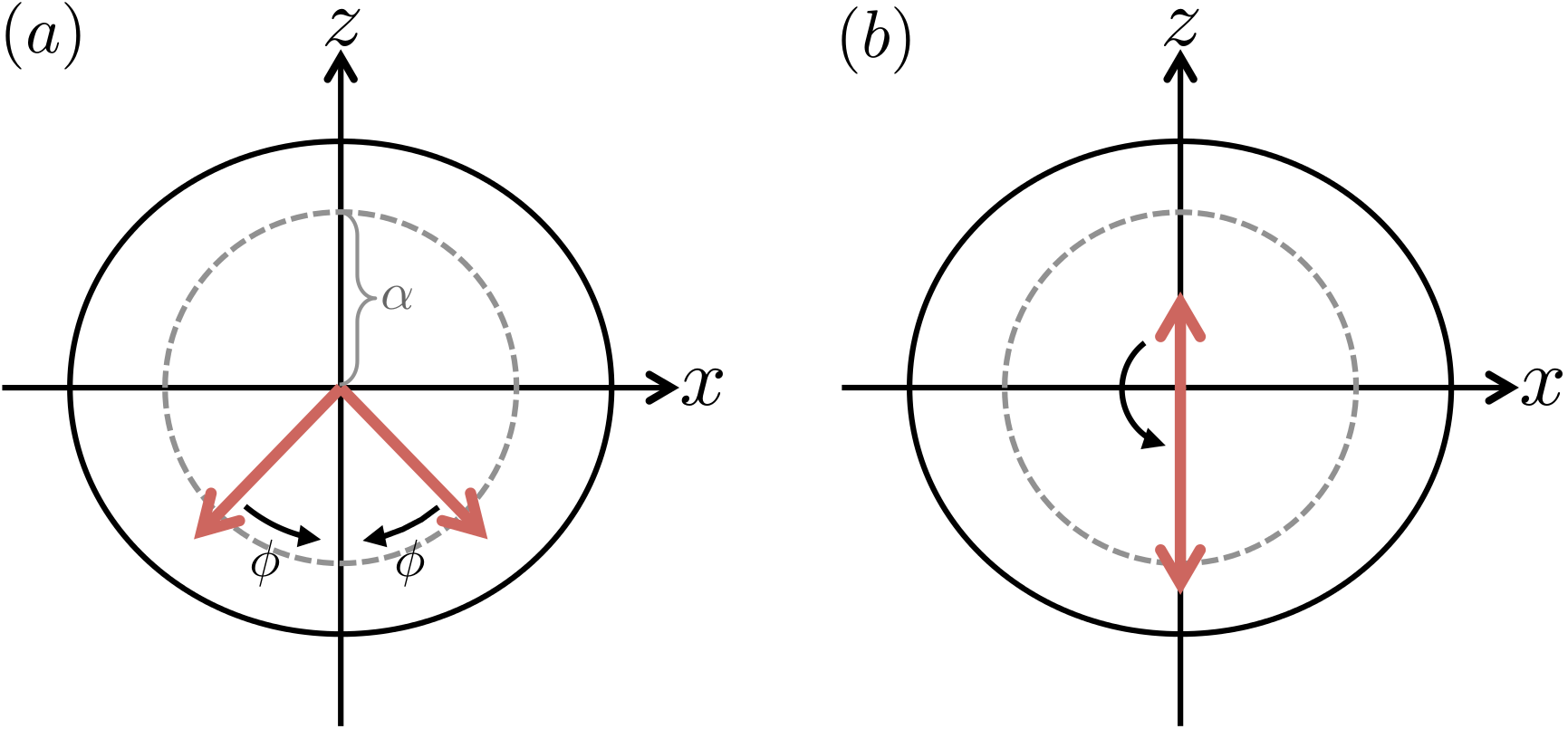}
\caption{$(a)$ Depiction of the conditional evolution of $U^x_{\pm}$ on the post-measurement state given a $x$-measurement outcome $|+\rangle$ (right) or $|-\rangle$ (left).  The grey dotted circle denotes the initial length of the Bloch vector, $\alpha$. $(b)$ Depiction of the conditional evolution of $U^z_{\pm}$ after a $z$-measurement.}\label{fig:rotation}
\end{figure}

Since it is a projective measurement in the auxiliary's $x$-basis that provides the most reliable readout of which direction (clockwise or counter-clockwise) the auxiliary was rotated~\cite{Levitin95}, such an $x$-measurement maximizes the classical mutual information between the outcome and the $\sigma_{\vec m}$ eigenstate of the system.
As a result, this measurement also produces the greatest reduction in the entropy (uncertainty) of the system.
This suggests that for our feedback protocol to obtain the maximal entropy reduction we must perform the feedback conditional on the auxiliary's $\sigma_x$ eigenstates $|+\rangle$ and $|-\rangle$. We therefore choose the feedback unitary to be $U^{\vec m}_{\rm fb}=U_+^{\vec m}|+\rangle\langle+|+U_-^{\vec m}|-\rangle\langle-|$, where the unitary operators 
\begin{equation}\label{eq:Ufeed1}
\begin{split}
U^x_{\pm}&=e^{\mp i\phi\sigma_y/2},\quad \phi=\tan^{-1}\left(\frac{\lambda}{\alpha}\tan(\theta)\right), \\
U^z_{+}&=e^{-i\pi\sigma_y/2} \quad{\rm and}\quad U^z_{-} = I 
\end{split}
\end{equation}
generate rotations on the Bloch sphere chosen so that they lead to the maximum reduction in entropy after measurement and feedback for a given measurement strength $\theta$, and mixing constants $\alpha$ and $\lambda$.
Their effects on the post-measurement state of the system are illustrated in Fig.~\ref{fig:rotation}, where we see that they each rotate the system so that the final state is always pointing down ($-{\hat z}$).

We now examine the entropy production and thermodynamic efficiency of unitary measurement-based feedback control for different choices of the measurement basis, either a $z$-measurement ($\vec{m} = {\hat z}$) or an $x$-measurement ($\vec{m} = {\hat x}$). 
Since the system is initially in a classical mixture of its $z$-states, the $z$-measurement is classical; that is, the change in the state of the system conditioned on the measurement outcomes can be described by Bayesian inference, and as such does not disturb the system. By contrast, the $x$-measurement has uniquely quantum features, because the density matrix has off-diagonal elements in the $x$-basis. In fact, the $x$-measurement leads to a greater reduction in the system's entropy (higher purity) after measurement and feedback, since the final length of the system's Bloch vector using the $x$-measurement,  
\begin{equation}
\gamma_x=\sqrt{\alpha^2\cos^2\theta+\lambda^2 \sin^2\theta}, 
\end{equation}
always exceeds that for the $z$-measurement, 
\begin{equation}\label{eq:gammaz}
\gamma_z=\lambda \sin\theta\le \gamma_x,
\end{equation}
except when $\theta=\pi/2$, when the measurement is a so-called ``infinite strength'' measurement~\cite{FJ}. 

\begin{figure}[tb]
\includegraphics[width=1\hsize]{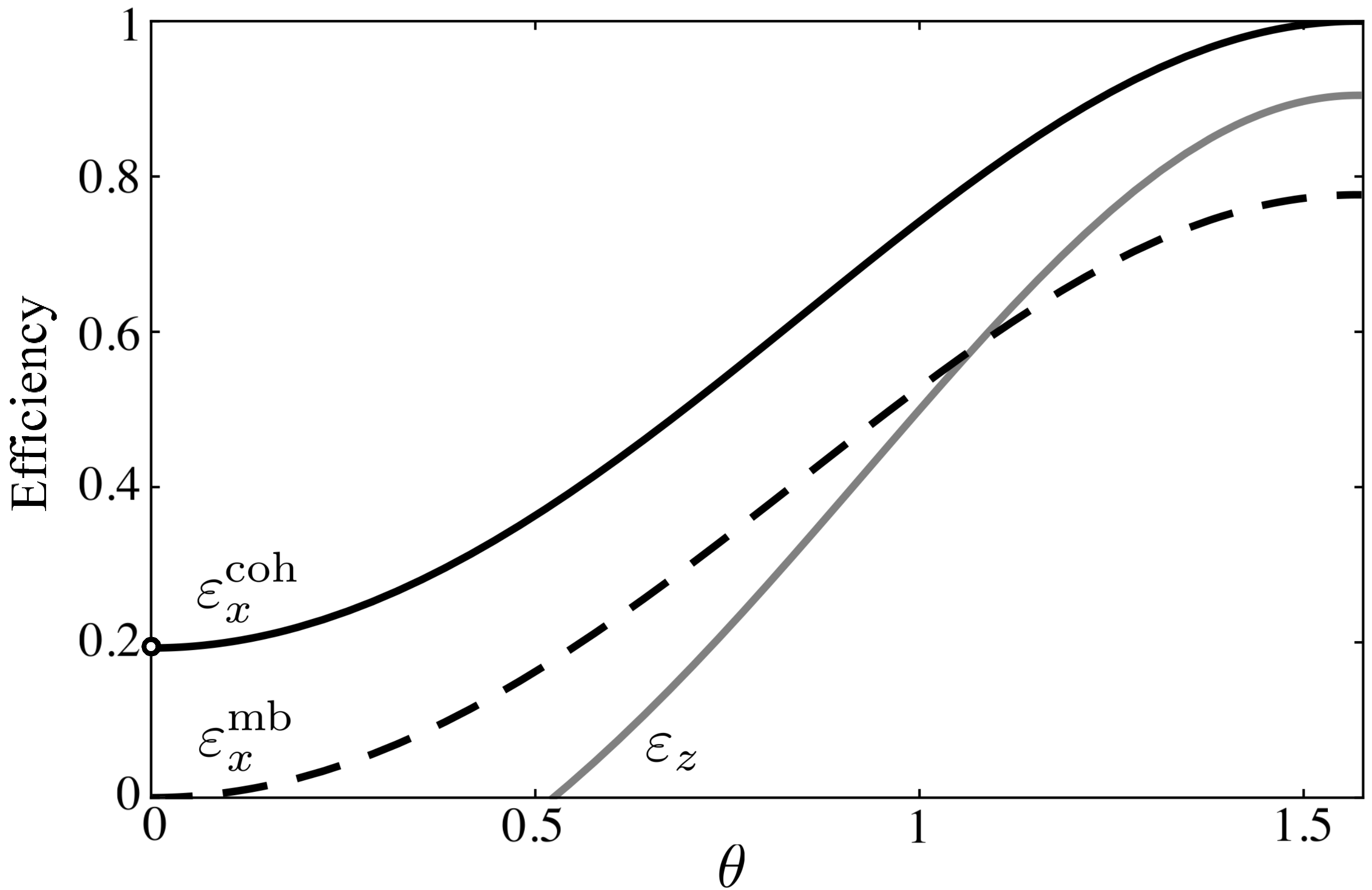}
\caption{Plot of the efficiency of the measurement-based feedback protocol using and explicit $x$-measurement, $\varepsilon_x^{\rm mb}$ (dashed) [Eq.~(\ref{eq:effMbx})], the same measurement-based protocol performed coherently, $\varepsilon^{\rm coh}_x$ (solid) [Eq.~(\ref{eq:effCohx})], and the measurement-based protocol that uses a $z$-measurement, $\varepsilon_z$ (grey) [Eq.~(\ref{eq:effz})] as functions of the strength parameter $\theta$ with $\alpha=0.4$ and $\lambda=0.8$. Note that at $\theta = 0$ the efficiency is not defined, since there is no measurement, and as a result the protocol does not change the entropy of the system.}
\label{fig:effPlot}
\end{figure}

We have determined the efficiency of measurement-based feedback control with an optimal reset [Eqs.~(\ref{eq:eff2}) and (\ref{eq:effMb})] for both the $x$- and $z$-measurements when the measurement is explicit, in which we decohere the auxiliary after the measurement (and the measurement results are classical numbers processable by a classical device); and when the auxiliary is not decohered, so that the measurement-based feedback procedure is performed coherently. For the $x$-measurement, the efficiency of explicit measurement-based feedback is 
\begin{equation}\label{eq:effMbx}
\varepsilon_x^{\rm mb}=\frac{h(\alpha)-h(\gamma_x)}{\ln 2-h(\lambda)},
\end{equation}
and for coherent measurement-based feedback it is 
\begin{equation}\label{eq:effCohx}
\varepsilon^{\rm coh}_x= \frac{h(\alpha)-h(\gamma_x)}{h(\alpha \lambda /\gamma_x)-h(\lambda)}.
\end{equation} 
By contrast, the $z$-measurement efficiency
\begin{equation}\label{eq:effz}
\varepsilon_z=\frac{h(\alpha)-h(\gamma_z)}{h(\alpha\gamma_z)-h(\lambda)} 
\end{equation}
does not depend on whether the feedback is performed coherently, due to its classical nature. In Fig.~\ref{fig:effPlot}, we plot as a function of $\theta$ the three efficiencies $\varepsilon_x^{\rm coh}$, $\varepsilon_x^{\rm mb}$, and $\varepsilon_z$ for representative values of $\alpha$ and $\lambda$, though we have verified that the conclusions are qualitatively similar for other values.
We observe that the coherent  $x$-measurement is always more efficient than the $z$-measurement, $\varepsilon_x^{\rm coh}> \varepsilon_z$. 
This is unexpected because usually thermodynamic efficiency is not improved by the coherence properties of quantum systems, although we have already seen how the decoherence step reduces the efficiency, which would suggest a lower classical z-measurement efficiency.
What is surprising is that when the $x$-measurement is explicit, and thus involves decoherence, it still outperforms the classical measurement when the measurements are weak ($\theta\lesssim 1.0$ for the values in Fig.~\ref{fig:effPlot}).  

One of the primary results of our analysis is that the ultimate source of inefficiency is the residual correlation between the system and auxiliary that remains after the feedback has been completed. One way to think about this is that the auxiliary obtains information about the system (the correlations) that it can then use to extract entropy. It would then seem natural to conclude that any unused correlations would reduce the efficiency, because they represent an unexploited resource.
But it is not quite that simple. Depending on the measurement basis and the rotation angle $\theta$, not all the correlations may be available to reduce the entropy of the system. Further, the correlations may actually be increased during the feedback step, so that the simple idea that correlations are ``used up'' in reducing the entropy of the system, while true classically, is not necessarily valid for quantum systems.

\begin{figure}
\includegraphics[width=1\hsize]{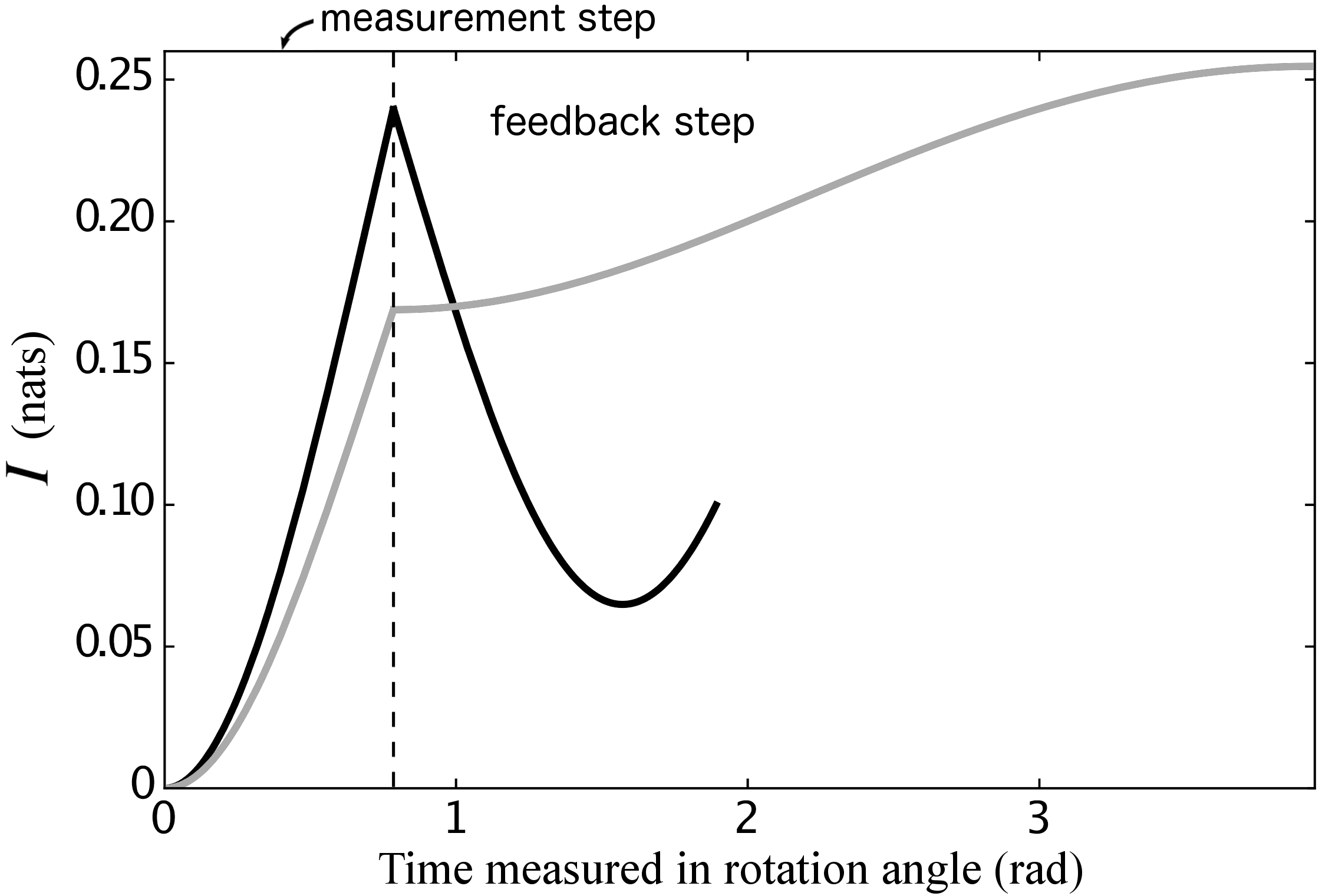} 
\caption{Time evolution of the mutual information for the $x$-measurement (black) and $z$-measurement during the course of measurement and feedback for a measurement strength $\theta=\pi/4\approx 0.78$, with an $x$-measurement feedback rotation of $\phi\approx1.89$ [Eq.~(\ref{eq:Ufeed1})].}\label{fig:infotime}
\end{figure}

To illustrate the above statements and to understand a little more why the quantum $x$-measurement performs better than the classical $z$-measurement, we consider the evolution of the mutual information (the correlations) over the course of the measurement and feedback steps. To do this we consider that the unitary $U_{\rm meas}$ is implemented by a Hamiltonian ($V_{\vec m}$), and follow the evolution of the mutual information as the system evolves under this Hamiltonian. We then continue this evolution under the Hamiltonian that generates $U^{\vec m}_{\pm}$ [Eq.~(\ref{eq:Ufeed1})]. We plot the mutual information as a function of time, using as our measure for time the angle through which the auxiliary is rotated during the measurement step, and the angle through which the system is rotated during the feedback step. Since the system must be rotated through a larger angle during the feedback step of the $z$-measurement ($\pi$ rotation) than the $x$-measurement ($\phi$ rotation), the $x$-measurement takes less time (for a constant rotation speed); yet another advantage of the quantum protocol. In Fig.~\ref{fig:infotime}, we plot the mutual information for both the $z$- and $x$-measurement protocols, for the case in which the measurement strength is $\theta=\pi/4$.
We see that during the measurement period the information increases, but the $x$-measurement gathers more information. The most surprising effect occurs during the coherent  feedback step after $\theta=\pi/4$: for these parameter values the quantum $x$-measurement displays a decrease in the information as one would expect, by contrast in the classical case the information actually increases.
The underlying reason is that due to the coherent nature of the interaction, coupled with the fact that the auxiliary states that are correlated with the system are not orthogonal, the entropy of the auxiliary continues to grow during the feedback step, reducing the efficiency of the control process. This is in stark contrast to feedback in classical systems where it is typically assumed the auxiliary (or memory) is fixed during the feedback step~\cite{Horowitz2013, Sagawa2013b}.

We have seen that maintaining coherences during a measurement-based feedback protocol can increase the energetic efficiency. More surprisingly, we have observed that measuring the system in a basis in which the density matrix is not diagonal, and thus has coherences, leads both to a greater reduction in entropy [Eq.~(\ref{eq:gammaz})] and a greater energy efficiency: the quantum protocol cools more effectively and more efficiently. 

\section{Conclusion}

In conclusion, we have developed and analyzed an energetic efficiency for quantum feedback control, and applied it to show that quantum measurements can increase the efficiency of  measurement-based feedback. Our framework can also be used to compare the thermodynamic efficiency of the feedback methods considered here to a variety of cooling scenarios~\cite{Ganes2008, Allahverdyan2011}. Further, while our analysis focused on the energetics of a single feedback step, repeated feedback is more commonly encountered in experiment~\cite{Koch10, Sayrin11, Vijay12}. Developing our framework for this situation is an interesting topic for future work. Particularly interesting is the possibility of erasing the results of a sequence of measurements in a combined process, leading to increased efficiency. 

\appendix

\acknowledgements
We are grateful to Massimiliano Esposito for his comments on an early draft of this manuscript. This work was partially supported by the NSF under Project Nos. PHY-1005571 and PHY-1212413, and by the ARO MURI grant W911NF-11-1-0268. 
 
\bibliography{report,Feedback.bib,PhysicsTexts} 

\end{document}